\documentstyle[12pt,aaspp4]{article}

\newcommand{\msun} {{{\rm\ M}_{\sun}}}

\begin{document}

\title{Updated Information on the Local Group}
\author{Sidney van den Bergh}
\affil{
Dominion Astrophysical Observatory \\
National Research Council of Canada \\ 
5071 West Saanich Road \\
Victoria, British Columbia, V8X 4M6 \\
Canada \\
e-mail: Sidney.vandenBergh@hia.nrc.ca }

\begin{abstract}

The present note updates the information published in my recent 
monograph on \underline{The Galaxies of the Local Group}.  Highlights include 
(1) the addition of the newly discovered Cetus dwarf spheroidal 
as a certain member of the Local Group, (2) an improved distance 
for SagDIG, which now places this object very close to the edge 
of the Local Group zero-velocity surface, (3) more information 
on the evolutionary histories of some individual Local Group 
members, and (4) improved distance determinations to, and 
luminosities for, a number of Local Group members.  These data 
increase the number of certain (or probable) Local Group members 
to 36.  The spatial distribution of these galaxies supports 
Hubble's claim that the Local Group ``is isolated in the general 
field.'' Presently available evidence suggests that star formation 
continued much longer in many dwarf spheroidals than it did in 
the main body of the Galactic halo.  It is suggested that ``young''
globular clusters, such as Ruprecht 106, might have formed in now 
defunct dwarf spheroidals.  Assuming SagDIG, which is the most 
remote Local Group galaxy, to lie on, or just inside, the 
zero-velocity surface of the Local Group yields a dynamical age 
$\gtrsim 17.9 \pm 2.7$~Gyr. 

\end{abstract}

\keywords{ galaxies:  clusters:  general --- Local Group}

\section{Introduction}

The study of individual nearby galaxies that belong to 
the Local Group is presently one of the most active areas of 
extragalactic research.  As a result the recent reviews by 
van den Bergh (1999a, 2000) are already slightly out of date.  
The purpose of the present paper is to provide an updated report 
on the status of the most recent research on individual Local 
Group members, and on the structure of the Local Group itself.

\section{Information on Individual Galaxies}

The individual galaxies for which new information has become 
available are listed below in approximate order of decreasing 
luminosity.

\subsection{The Andromeda galaxy (M31 = NGC 224)}

Kormendy \& Bender (1999) have used spectroscopy with high angular 
resolution to study the binary nucleus of M31.  Their observations 
support Tremaine's model in which P$_1$ is the brightest part of a 
single eccentric disk, where stars linger while at the apocenters 
of their orbits around P$_2$.  The latter object is found to contain 
a $3.3 \times 10^7 \msun$ black hole.  
Barmby et al. (1999) have published a 
catalog of 435 probable globular clusters in M31, of which 330 
have UBV photometry, and 158 have been observed spectroscopically.  
These authors find that the metal-poor clusters have larger 
projected distances from the galaxy center, and show lower rotation, 
than do the metal-rich clusters.  They estimate
that M31 contains $450 \pm 100$ globulars, from which the specific globular 
cluster frequency ${\rm S} = 0.9 \pm 0.2$.  Hamilton \& Fesen (1999) have used 
the Hubble Space Telescope to image absorption by the remnant 
of the supernova S Andromedae 
(SN 1885A), which is observed to have a diameter of 
$0^{\prime\prime}.55 \pm 0^{\prime\prime}.15$, 
in the light of Fe II.  From their observations these authors 
conclude that this remnant contains 
between 0.1 and 1.0 $\msun$ of iron.

\subsection{The Milky Way system}

	Many years ago Eggen \& Sandage (1959) suggested that RR 
Lyrae and Groombridge 1830 (HR 4550) belong to a small physical 
group of five stars that are traversing the solar neighborhood at 
high velocity.  The view that such clumps of high-velocity stars 
are real is supported by observations with the HIPPARCOS satellite 
(Helmi et al. 1999).  From space motions of a nearly complete 
sample of nearby high-velocity halo stars these authors conclude 
that seven objects are members of a single debris stream.  From their 
data Helmi et al. conclude that
$\sim$8\% of all metal-poor stars outside the solar radius represent the 
remnants of a single disintegrated dwarf galaxy.  However, this 
estimate may turn out to be too high if a single narrow debris stream 
presently happens to be passing through the solar neighborhood.

	Figer et al. (1999) have determined the slope of the mass 
spectrum of in the Arches and Quintuplet clusters, which are located 
near the Galactic center.   They find that the mass spectra of these 
very young clusters have a slope $\Gamma = -0.65$, 
which is less steep than that for young clusters elsewhere in 
the Galaxy which typically have $\Gamma \sim -1.4$.  
Each cluster has a mass of  
$\sim 1 \times 10^4 \msun$, 
which are among the highest known in the Galaxy.  Taken 
at face value these results suggest that the region near the Galactic 
center is particularly prone to forming very massive open clusters.  
From near-infrared echelle spectra Carr, Sellgren \& Balachandran (1999) 
find that [Fe/H] = -0.02 $\pm$ 0.13, i.e. nearly solar, for the Galactic 
center supergiant IRS 7.

\subsection{The Triangulum galaxy (M33 = NGC 598)}

	Corbelli \& Salucci (1999) have measured the rotation curve 
of M33 out to a distance of 16 kpc (13 disk scale-lengths) and find 
that the rotation curve rises out to the last measured point.  This 
result implies a dark halo mass of 
$\gtrsim 5 \times 10^{10} \msun$.  Beyond 3 kpc the 
gravitational potential is dominated by a dark halo with a density 
that decreases radially as R$^{-1.3}$.  From an unbiased sample of 60 
clusters Chandar et al. (1999a,b) find that cluster formation in 
M 33 has been continuous over the last 10 Gyr, i.e. unlike the LMC, 
M 33 did not have a gap in its cluster formation history.  Young 
clusters in M 33 have masses in the range 
$6 \times 10^2 \msun$ to $2 \times 10^4 \msun$, 
which is smaller than those of the old clusters which typically have 
masses of a few $\times 10^5 \msun$.  Gordon et al. (1999) have doubled the 
sample of supernova remnants that are known in the Triangulum galaxy.  
Among their sample of 53 SNRs they find no evidence for a strong 
correlation between surface brightness and diameter.  Many of these 
remnants are found to be associated with, or embedded in, H II regions.  
This (not unexpectedly) suggests that the majority of these remnants 
were produced by supernovae of Type II.

\subsection{The Large Magellanic Cloud (LMC)}

	Detached eclipsing variables are powerful tools for the 
determination of extragalactic distances.  Recently Nelson et al. 
(1999) have redetermined the reddening of the eclipsing variable 
HV 2274 and find E(B-V)~$ = 0.083 \pm 0.006$, which is significantly 
lower than previously published values.  From their new reddening 
Nelson et al. derive a distance modulus (m-M)$_0 = 18.40 \pm 0.07$.  
Gibson  (1999) has reviewed recent distance determinations for the 
LMC, which range from (m-M)$_0 = 18.20$ to 18.75.  This large spread 
shows that significant unappreciated sources of systematic error 
still exist in modern determinations of the distance to the Clouds 
of Magellan.

Sakai, Zaritsky \& Kennicutt (1999) have recently used the magnitude  
level of the tip of the LMC red giant branch to derive a distance 
modulus (m-M)$_0  = 18.59 \pm 0.09$ (random) $\pm 0.16$ 
(systematic), which agrees well with previous determinations via Cepheid 
variables.

	Dolphin (1999b) has studied the star formation history in 
two fields in the LMC.  His results, which are summarized in Table 1, 
appear to show (1) a steady  rise of the metallicity index [Fe/H] 
with time, and (2) that the rate of star 
formation between 2.5 and 7 Gyr ago was an order of magnitude 
lower than it has  
been during the most recent 2--3 Gyr period. Since the two fields 
studied by 
Dolphin are separated by $3^{\circ}.0$ (2.6~kpc) his data refer to global 
star formation rates.  
Taken at face value Dolphin's results appear to weaken the 
previous conclusion that the rate of cluster formation in the 
Large Cloud increased more rapidly  $\sim$3 Gyr ago than did the 
rate of star formation.  However, Holtzman et al.  (1999) reach 
a very different conclusion from \underline{Hubble Space Telescope} 
observations of two fields in the outer disk of the LMC.  They conclude 
that there was no gap in the age distribution of these stars.  Clearly 
we are still far from understanding the evolutionary history of the 
disk component of the LMC.   From new HST color- magnitude diagrams 
Johnson et al. (1999) find that the age difference between the Large 
Cloud clusters NGC 1466, NGC 2257 and Hodge 11, on the one hand, and 
the Galactic clusters M 92 and M 3 on the other, is less than 1.5 Gyr.  
It is presently not understood why globular cluster formation occurred 
simultaneously in the LMC, the main body of the Galactic halo, 
and in the outer Galactic halo  (NGC 2419).

	From the relatively large number of red clump stars in the 
LMC Bar Holtzman et al. conclude that the stellar population in the 
Bar is older than that in the outer fields of the Large Cloud.  This 
conflicts with some previous work that had suggested that the LMC Bar 
might be a relatively recent feature.  From their integrated spectra 
Dutra et al. (1999) conclude that NGC 1928 ([Fe/H]~$ = -1.2$) and 
NGC 1939 ([Fe/H]~$ = -2.0$) may be globular clusters that had not 
previously been recognized as such.  If confirmed this would 
increase the number of LMC globular clusters from 13 to 15.  A 
catalog of X-ray sources in the LMC has been published by Haberl \& 
Pietsch (1999). These authors give likely identifications for 144 
sources.  Of these objects 46 (32\%) appear to be associated with 
supernova remnants, 17 (12\%) are X-ray binaries, and nine (6\%) are 
``supersoft'' sources. The majority of unidentified sources are 
probably associated with background galaxies or foreground stars.  
Demers \& Battinelli (1999) have surveyed the periphery of the LMC 
for young blue stars that might be associated with the Bridge 
linking the Large and the Small Clouds.  Few such stars are found 
suggesting that the Bridge does not extend deep into the LMC.  
Liebert (1999) has pointed out that the hot blue star found in the 
LMC cluster NGC 1818 has the wrong luminosity and radius to be a 
luminous white dwarf.

	Murali (1999) finds that the motion of the Magellanic 
Stream through  ambient gas can strongly heat the Stream clouds, 
driving mass loss and causing  evaporation.  Survival of the 
stream for 500 Myr sets an upper limit $< 10^{-5} {\rm\ cm}^{-3}$ 
for the  Galactic halo gas through which the Stream in orbiting.

\subsection{The Small Magellanic Cloud}

	Most previous investigators have concluded that cluster 
formation in the Small Cloud has proceeded at a more-or-less 
uniform rate.  However new cluster age determinations by Rich 
et al. (1999) now suggest that cluster formation in the SMC 
may have been enhanced during bursts that occurred 2 Gyr and 
8 Gyr ago.  More, and more accurate, age determinations for 
Small Cloud clusters will be required to strengthen this conclusion.  
An unbiased survey of 93 star clusters in a 2.4 square degree 
area of the SMC (Pieterzy\'{n}ski \& Udalski 1999) shows an age 
distribution that is very strongly biased towards young clusters, 
with only 3\% of the clusters having ages $> 1$~Gyr.  The fact that 
60\% of all SMC clusters are younger than 100 Myr should probably 
be interpreted as evidence for short cluster life-times, rather 
than as evidence for a recent burst of cluster formation.

Bica \& Dutra (1999) have published an updated census of SMC  
clusters, and of clusters in the Bridge between the LMC and SMC, 
that is based on the  OGLE survey.  Their paper contains a map 
of the distribution of SMC  clusters, which shows a strong 
concentration in the Small Cloud ``Bar'' and a lesser  concentration 
of clusters in the Wing of the SMC.

	Cold atomic hydrogen has been detected in the Bridge between 
the Magellanic Clouds by Kobulnicky \& Dickey (1999).  The 
early-type stars observed in the Bridge could have formed in 
these cold clouds.  These objects therefore need not have migrated 
from the main body of the SMC.

	Rolleston et al. (1999) have compared the abundances of 
three early-type stars in the Bridge between the LMC and the SMC 
with those observed in ``normal'' B-type stars near the Sun.  They 
found an average metal deficiency [m/H]~$ = -1.05 \pm 0.12$ for these 
objects.  Surprisingly this value falls well below the present 
(Luck et al. 1998) SMC metallicity [Fe/H]~$ = -0.74$.  In fact it 
lies between that of the globular cluster NGC 121 ([Fe/H]~$ = -1.19$), 
which has an age of  $\sim$12 Gyr, and that of the old open cluster 
L~1 ([Fe/H]~$ = -1.01$) that has an age of  $\sim$10 Gyr.  Taken at face 
value this result suggests that the material in the Bridge was 
tidally detached from the Small Cloud  $\sim$10 Gyr ago.  However, 
Putman (1999) has argued that the Bridge was formed only 0.2 Gyr 
ago, when the LMC and SMC had a very close encounter with a 
minimum separation of only
$\sim$7 kpc.  Alternatively it might be hypothesized that the 
metallicity of gas in the Bridge was lowered by swept-up 
pristine inter-galactic gas.  The metallicity of NGC 330, 
which is the most luminous young SMC cluster, remains 
controversial.  Hill (1999) finds 
$\langle[{\rm Fe/H}]\rangle = -0.82 \pm 0.11$ for 
six cool cluster supergiants, compared to 
$\langle[{\rm Fe/H}]\rangle = -0.69 \pm 0.10$
for six cool SMC field stars.  Finally Gonzalez \& Wallerstein (1999) 
obtain $\langle[{\rm Fe/H}]\rangle = -0.94 \pm 0.02$ for seven stars in NGC 330.

	The ninth nova to be discovered in the SMC was found 
near the optical  center of this galaxy by Glicenstein (1999) 
at 
$\alpha = 0^{\rm h}\ 59^{\rm m}\ 23.^{\rm s} 0,\ 
\delta = -73^{\circ}\ 07^{\prime}\ 56^{\prime\prime}$ (equinox 2000).

\subsection{The spheroidal{\protect \footnotemark[1]} galaxy NGC 205}

	A nova has been discovered in NGC 205 by Johnson \& Modjaz (1999).

\footnotetext[1]{The galaxies NGC 147, NGC 185, and NGC 205 belong to the same 
morphological family as do the dwarf spheroidals, such as Sculptor and 
Fornax.  However, their relatively high luminosity makes the term dwarf 
spheroidal seem inappropriate.  I therefore call such objects ``spheroidals''.  
It seems likely that such spheroidals are lower luminosity examples of the 
form family that de Vaucouleurs (1959) refers to as ``lenticulars''.}

\subsection{The starburst galaxy IC 10}

	From observations in the far infrared Bolatto et al. (1999) 
find that dust in the mild starburst galaxy IC 10 appears to deficient 
in small grains.  It seems likely that such grains were destroyed by 
intense UV radiation in the neighborhood of the hot luminous stars 
that were formed during the recent burst of star formation in this galaxy.

\subsection{The spheroidal galaxy NGC 185}

	The history of star formation in NGC 185 has been studied by 
Mart\'{\i}nez-Delgado, Aparicio \& Gallart (1999), who find that the bulk 
of the star formation in this galaxy took place at early times.  
Stars only formed near the center of this galaxy during the last  
$\sim$1 Gyr.  Most of the young blue objects discovered by Baade (1951) 
turn out to be star clusters, rather than individual stars.  
Mart\'{\i}nez-Delgado et al. have also discovered a supernova remnant 
near the center of NGC 185.

\subsection{The dwarf irregular IC 1613}

	\underline{Hubble Space Telescope} observations by Cole et al. (1999) 
show that the dominant stellar population in IC 1613 has an age 
of  $\sim$7 Gyr.  From its Hess diagram these authors find that the 
evolutionary history of IC 1613 may have been similar to that of 
the Pegasus dwarf (DDO 216).  Both of these objects are classified 
Ir~V on the DDO system.  Antonello et al. (1999) have found five 
Cepheids of Population II in IC 1613, thus providing \underline{prima} 
\underline{facie}
evidence for the existence of a very old stellar population 
component in this galaxy.  King, Modjaz \& Li (1999) have observed 
what is believed to be the first nova ever observed in IC 1613.

\subsection{The Wolf-Lundmark-Melotte system (DDO 221)}

	A recent color-magnitude diagram obtained with the 
\underline{Hubble Space Telescope} 
(Dolphin 1999a) shows that about half of all star 
formation in the WLM galaxy occurred during a burst that began  
$\sim$13 Gyr ago.  During the course of this burst the metallicity 
increased from [Fe/H]~$\sim$$-2.2$ to [Fe/H]~$= -1.3$.  From the apparent 
absence of a horizontal branch population Dolphin places an upper 
limit of  $\sim$$20 \msun$ per Myr on the rate of star formation between 
12 Gyr and 15 Gyr ago.  Between 2.5 and 9 Gyr ago the average rate 
of star formation was $100 - 200 \msun$ per Myr.

	The WLM system contains a single globular cluster, for 
which Hodge et al. (1999) find M$_{\rm V} = -8.8$, [Fe/H]~$ = -1.5$ 
and an age of  $\sim$15 Gyr.  It is of interest to note that the lone 
globular in this dwarf galaxy has a luminosity that falls slightly 
above that of the mean for all globular clusters.  The apparent 
absence of faint low-mass globulars is of particular interest 
because destruction of such objects probably can not be attributed 
to the weak tidal forces of the WLM dwarf galaxy.

\subsection{The disintegrating Sagittarius galaxy}

From observations of tidal debris of the Sagittarius system 
Johnston et al. (1999) conclude that this object has orbited the 
Galaxy for at least 1 Gyr, and that the mass of this galaxy has 
decreased by a factor of 2--3 during this period.  The orbits of 
Sgr are found to have Galactocentric distances that oscillate between
$\sim$13 kpc and  $\sim$41 kpc and periods in the range 550--750 Myr.  
Its most recent perigalacticon occurred  50 Myr ago.  [Jiang \& 
Binney (1999) find that the Sagittarius dwarf might have started 
its infall from a distance greater than 200 kpc if its initial mass 
was as great as  $\sim$10$^{11} \msun$.]  Both the orbit of Sgr, 
and the Galactic 
potential field, could be constrained by improved proper motion 
observations of the stellar debris associated with this object.  
Burton \& Lockman  (1999) have found no neutral hydrogen gas 
associated with the Sagittarius dwarf.

\subsection{The Fornax dwarf spheroidal}

	Buonanno et al. (1999) have recently used the 
\underline{Hubble Space Telescope} 
to obtain a color-magnitude diagram for the globular 
cluster Fornax No. 4.  Whereas the clusters Fornax 1, 2, 3 and 5 
have horizontal branches that extend over a wide range of colors 
(and include RR Lyrae variables), Fornax 4 is seen to have a red 
horizontal branch.  Fornax 4 is  $\sim$3 Gyr younger than the other 
Fornax globulars.  Buonanno et al. draw attention to the fact that 
the color-magnitude diagram of Fornax 4 exhibits a strong resemblance 
to that of the ``young'' Galactic globular cluster Ruprecht 106 
(Fusi Pecci et al. 1995, and references therein).  This observation 
raises two questions:  (1) What is it about dwarf spheroidals that 
allows them to form globular clusters such as Fornax 4 and 
Terzan 7 (which is associated with the Sagittarius dwarf) long 
after the formation of globulars ceased in the main body of the 
Galactic halo, and (2) could it be that ``young'' outer halo globulars 
such as Ruprecht 106 were originally formed in now defunct dwarf 
spheroidals? 

\subsection{The Sagittarius dwarf irregular galaxy (SagDIG)}

	On the basis of a very uncertain distance of 1.4 Mpc 
SagDIG (UKS 1927-177~$=$~UGA 438) had previously been regarded 
as a possible Local Group member.  New photometry by Lee \& Kim 
(1999) yields a distance (based on the tip of the red giant branch) 
of only $1.18 \pm 0.10$ Mpc, and an improved integrated luminosity 
M$_{\rm V} = -11.97$, which makes this object slightly more luminous that 
Leo A (M$_{\rm V} = -11.5$).  A similar distance of $1.06 \pm 0.10$ Mpc has 
recently been found by Karachentsev, Aparicio \& Makarova (1999).  
If, following Courteau \& van den Bergh (1999), we assume that the 
barycenter of the Local Group is situated at a distance of 454 kpc 
towards $\ell = 121.^{\circ}7, {\rm b} = -21.^{\circ}3$, 
then SagDIG is located at a distance 
of $1.29 \pm 0.09$ Mpc from the LG barycenter.  This makes SagDIG the 
most remote object suspected of Local Group membership.  Its distance 
is marginally greater than the radius of the zero-velocity surface 
of the Local Group, for which Courteau \& van den Bergh (1999) find 
a value of R$_0 = 1.18 \pm 0.15$ Mpc.

\subsection{The Leo I (Regulus system)}

	From HST observations Gallart et al. (1999) conclude that 
70\% -- 80\% of the star forming activity in Leo I took place between 
7 Gyr and 1 Gyr ago.  There is little or no evidence for the 
presence of stars with ages  $> 10$ Gyr.  About 1 Gyr ago the rate 
of star formation appears to have dropped abruptly to a near-negligible 
level.  However, some very low-level star formation 
may have continued until  $\sim$300 Myr ago.

\subsection{The M31 companion And II}

	From spectra of seven stars in And II, that were obtained 
with the Keck telescope, C\^{o}t\'{e} et al. (1999) have found a mean 
velocity V$_{\rm r} = -188 \pm 3 {\rm\ km\ s}^{-1}$, 
and a velocity dispersion of $9.3 \pm 2.6 {\rm\ km\ s}^{-1}$.  
From these data they obtain a mass-to-light 
ratio M/L$_{\rm V} = 21^{+14}_{-10}$ in solar units, 
i.e. this dwarf spheroidal appears 
to contain a significant amount of dark matter.  C\^{o}t\'{e}, Oke \& 
Cohen (1999) have also obtained Keck spectra of 42 red giants 
in And II, from which they find a mean metallicity 
$\langle[{\rm Fe/H}]\rangle = -1.47 \pm 0.19$, 
with a dispersion of $0.35 \pm 0.10$ dex.  
Da Costa et al. (1999) have studied the color-magnitude diagram 
of And II and find that the majority of stars have ages in the 
range 6 Gyr to 9 Gyr, although the presence of RR Lyrae variables 
and blue horizontal branch stars attests to the existence of a 
population component with an age $> 10$ Gyr.  And II differs from 
And I in that it does not exhibit a radial gradient in horizontal 
branch morphology. Furthermore the dispersion in abundance is 
considerably larger in And II than it was found to be in And I.  
These results show that these two dwarf spheroidal companions to 
M 31 must have had quite different evolutionary histories.  It 
would be interesting to know if there is a correlation between a 
radial horizontal branch gradient and the metallicity dispersion 
among dwarf spheroidal galaxies. On the basis of its 680 kpc 
distance Da Costa et al. conclude that And II is physically 
associated with M 31, rather than with M 33.

\subsection{The dwarf spheroidals And V, And VI and And VII}

Caldwell (1999) has derived accurate luminosities and surface 
brightness profiles for Andromeda V, VI and VII.  And V turns 
out to be fainter than previously believed, whereas And VI and 
And VII were found to be more luminous than previously thought.   
And V has a metallicity that lies above the average metallicity 
versus luminosity relation for Local Group dwarf galaxies.

\subsection{The Aquarius dwarf (AqrDIG = DDO 210)}

The membership of this dwarf irregular in the Local Group has now 
been firmly established by Lee et al. (1999), who derive a distance 
of $950 \pm 50$ kpc from the magnitude of the giants at the tip of the 
red giant branch.  The corresponding distance of this object from 
the barycenter of the Local Group is also 950 kpc.  This implies 
that DDO 210 is rather isolated in space.

\subsection{The recently discovered Cetus system}

	Whiting, Hau \& Irwin (1999) have searched the region 
with $\delta < +3^{\circ}$ for faint dwarfs that might be members of the 
Local Group.  Two objects were discovered during this program.  
One of these was the Antlia system, which lies just beyond the 
zero-velocity surface of the Local Group (van den Bergh 1999b). 
The other was a faint dwarf spheroidal galaxy in Cetus.  From the 
position of the tip of the giant branch of the Cetus system 
Whiting et al. derive a Galactocentric distance of $775 \pm 50$ kpc, 
and a distance of 615 kpc from the adopted center of the Local 
Group.  This places Cetus comfortably inside the 1.18 Mpc radius 
(Courteau \& van den Bergh 1999) of the Local Group zero-velocity 
surface.  It would be of great importance to obtain radial 
velocities for individual red giants in the Cetus dwarf.  This 
would enable one to determine the amount of dark matter in this 
galaxy.  Furthermore knowledge of the systemic velocity of Cetus 
would add weight to the Local Group mass determination.  No neutral 
hydrogen gas has been found to be associated with the Cetus dwarf.

\subsection{The Sculptor dwarf spheroidal}

	Majewski et al. (1999) find that the metallicity 
distribution in Sculptor appears to be bimodal with components 
having [Fe/H]~$ = -2.3$ and [Fe/H]~$ = -1.5$.  As is the case in many 
other Local Group galaxies the older metal-poor component appears 
more extended than the younger metal-rich component.  However, 
Hurley-Keller, Mateo \& Grebel (1999), although confirming the 
central concentration of red horizontal branch stars, do not 
find a radial age (or metallicity) gradient. 

\subsection{The Phoenix dwarf galaxy}

	Mart\'{\i}nez-Delgado, Gallart \& Aparicio (1999) find that 
this dIr/dSph galaxy has an inner component that contains young 
stars which is stretched in the east-west direction, and an outer 
component that is extended north-south, which is mainly populated 
by old stars.  The rate of star formation in the central region 
of this galaxy appears to have remained approximately constant 
over time.  St.-Germain et al. (1999) have found a cloud of 
 $\sim$10$^5 \msun$ of H I located just west of Phoenix, that has a radial 
velocity of $-23 {\rm\ km\ s}^{-1}$, which may be physically associated with 
this object.  Optical radial velocities of stars in Phoenix will 
be required to confirm this suspicion.

\subsection{The Ursa Minor dwarf }

	New \underline{Hubble Space Telescope} observations by Mighell \& 
Burke (1999) confirm that the UMi system had a simple 
evolutionary history with a single  $\sim$2 Gyr long burst of 
star formation occurring  $\sim$14 Gyr ago.

\section{DISCUSSION}

	New data by Armandroff et al. (1999), Caldwell (1999), 
Da Costa et al. (1999), Dolphin (1999a), Lee \& Kim (1990), 
Lee et al. (1999), and Whiting et al. (1999) are incorporated 
in Table 2, which lists classification types on the DDO system, 
luminosities, Galactocentric distances, and distances from the 
adopted barycenter, for all 36 presently known Local Group 
members.  Figure 1 shows a plot of the integral frequency of 
galaxy distances from the Local Group barycenter adopted by 
Courteau \& van den Bergh (1999).  To minimize the influence of 
observational selection effects galaxies with M$_{\rm V} > -10.0$ 
have not been plotted.  The figure shows that only one galaxy 
(SagDIG at R$_{\rm LG} = 1.29 \pm 0.09$ Mpc) lies marginally beyond the 
dynamically determined radius of the Local Group zero-velocity 
surface, for which Courteau \& van den Bergh (1999) determined 
R$_0 = 1.10 \pm 0.15$ Mpc.  From Eqn. (10) of Courteau \& van den 
Bergh  (1999) it is seen that the age of the Local Group can 
be determined from its mass M, and the radius of its 
zero-velocity surface.  Adopting a Local Group mass 
M~$ = (2.3 \pm 0.6) \times 10^{12} \msun$ and assuming that SagDIG, which is 
located at a distance of $1.29 \pm 0.09$ Mpc from the barycenter 
of the Local Group, lies on the Local Group zero-velocity 
surface one obtains a value of $17.9 \pm 2.7$ Gyr for the dynamical 
age of the Local Group.  This value compares favorably with 
the 12--16 Gyr age derived from evolutionary models of the most 
metal-poor stars, the $16 \pm 6$ Gyr age that Sneden et al. (1996) 
derive from the Th/Eu abundance ratio in CS~22893-052, and the 
10--20 Gyr ages that Cowan, Thielemann \& Truran (1991) have  
derived from cosmochronology.  If the Local Group zero-velocity 
surface, in fact, lies slightly beyond SagDIG, then the LG age 
calculated above becomes a lower limit.  The figure shows that 
71\% of all Local Group galaxies\footnotemark[2] are situated within
0.5 Mpc of the Local Group barycenter.  On the other hand only 
one (4\%) of the known Local Group galaxies brighter than M$_{\rm V} = -10$ 
are located beyond R$_{\rm LG} = 1.0$ Mpc.  The total luminosity of the 
Local Group is found to be M$_{\rm V} = -22.0$, of which only 0.5\% originates 
beyond R$_{\rm LG} = 0.5$ Mpc.  This strongly supports Hubble's (1936) claim 
that the Local Group ``is isolated in the general field''.  The only 
Local Group member, in addition to SagDIG, that is known to have 
R$_{\rm LG} > 1.0$ Mpc is the faint Tucanae system (M$_{\rm V} = -9.6$) 
at R$_{\rm LG} = 1.1$ Mpc.

\footnotetext[2]{In van den Bergh (2000) all galaxies within $\sim$1.5 Mpc were initially 
regarded as candidate Local Group members.  Subsequently it was found 
that the Local Group has a zero-velocity surface with a radius of 
$1.10 \pm 0.15$ Mpc, as  measured from its barycenter.  All galaxies within 1.25 Mpc 
of the barycenter  have therefore been regarded as possible, or probable, 
Local Group members.}

Among the most important Local Group problems that remain to be 
resolved are the following:

\begin{itemize}

\item Why does the luminosity distribution of Local Group 
galaxies contain at least an order of magnitude fewer faint 
dwarfs than theory (Klypin et al. 1999) predicts?

\end{itemize}

Is theory wrong, or have the majority of low-mass dwarfs formed 
no stars?  A deep search for stars associated with compact 
high-velocity clouds might answer this question.

\begin{itemize}

\item Why does the Large Magellanic Cloud contain very old globular 
clusters that have disk kinematics, while M 33 is embedded in an 
apparently younger globular cluster system that exhibits halo kinematics?

\item Why did the LMC form so few clusters with ages between 4 Gyr 
and 10 Gyr? Was there a similar hiatus in the rate of star formation, 
or did the fraction of all Large Cloud stars that ended up in 
clusters suddenly increase  $\sim$4 Gyr ago? 

\item Did any of the Local Group galaxies change their morphological 
type during the last  $\sim$10 Gyr?  It would be particularly important 
to establish if the Bar of the LMC is a relatively young feature.

\item Why did different Local Group dwarf galaxies have such different 
star forming histories, and how do such differences depend on environment?

\item It would be important to strengthen and confirm Freeman's (1999) 
conclusion that the metal-rich r$^{1/4}$ halo of M 31 resulted from 
violent relaxation after the merger of two massive ancestral 
galaxies, whereas the metal-poor outer halo of the Milky Way 
system was mainly built up via capture of numerous low-mass 
metal-poor objects. 

\end{itemize}

The new generation of powerful optical, radio and space telescopes, 
in conjunction with improved wide-field and infrared-sensitive 
detectors, should enable us to answer many of these questions in 
the first years of the next millennium. 

\begin{acknowledgements}
I am indebted to Alan Whiting for making his data on the Cetus 
system available before publication.
\end{acknowledgements}

\begin{deluxetable}{rrr}
\tablecaption{STAR FORMATION HISTORY OF THE LARGE CLOUD\tablenotemark{a}}
\tablehead{ \colhead{Age (Gyr)} & \colhead{SFR ($\msun/$Myr)} & 
\colhead{[Fe/H] \tablenotemark{b} } }
\startdata
0.0 - 1.0 & 350 $\pm$ 20 & -0.38\nl

1.0 - 2.5 & 269 $\pm$ 40 & -0.54\nl

2.5 - 5.0 &  20 $\pm$ 30 & -0.83\nl

5.0 - 7.0 &  30 $\pm$ 40 & -1.06\nl

7.0 - 9.0 & 240 $\pm$ 70 & -1.32\nl

9.0 -12.0 & 440 $\pm$ 60 & -1.63\nl
\enddata
\tablenotetext{a}{Data from Dolphin (1999b)}
\tablenotetext{b}{These metallicity values are uncertain by  $\sim$0.1 dex}
\end{deluxetable}

\begin{deluxetable}{lllrll}
\tablecaption{DATA ON LOCAL GROUP GALAXIES\tablenotemark{a}}
\footnotesize
\tablehead{
\colhead{Name} & \colhead{Alias} & \colhead{Type} & \colhead{M$_{\rm V}$}  & \colhead{D (Galaxy)} & \colhead{D (Local Group)} }
\startdata
WLM & DDO 221 & Ir IV-V & -14.4 & 0.95 Mpc & 0.81 Mpc \nl
IC 10 & UGC 192 & Ir IV: & -16.3 & 0.66 & 0.27 \nl
Cetus &  & dSph & -10.1 & 0.78 & 0.62 \nl
NGC 147 & UGC 326 & Sph & -15.1 & 0.66 & 0.22  \nl
And III & A0032+36 & dSph & -10.2 & 0.76 & 0.31 \nl
NGC 185 & UGC 396 & Sph & -15.6 & 0.66 & 0.22 \nl
NGC 205 & M 110 & Sph & -16.4 & 0.76 & 0.31 \nl
M 32 & NGC 221 & E2 & -16.5 & 0.76 & 0.31 \nl
M 31 & NGC 224 & Sb I-II & -21.2 & 0.76 & 0.30 \nl
And I & A0043+37 & dSph & -11.8 & 0.81 & 0.36 \nl
SMC &  & Ir IV/IV-V & -17.1 & 0.06 & 0.48 \nl
Sculptor &  & dSph & -  9.8 & 0.09 & 0.44 \nl
Pisces & LGS 3 & dIr/dSph & -10.4 & 0.81 & 0.42 \nl
IC 1613 &  & Ir V & -15.3 & 0.72 & 0.47 \nl
And V &  & dSph & -  9.1 & 0.81 & 0.37 \nl
And II &  & dSph & -11.8 & 0.68 & 0.26 \nl
M 33 & NGC 598 & Sc II-III & -18.9 & 0.79 & 0.37 \nl
Phoenix &  & dIr/dSph & -  9.8 & 0.40 & 0.59 \nl
Fornax &  & dSph & -13.1 & 0.14 & 0.45 \nl
LMC &  & Ir III-IV & -18.5 & 0.05 & 0.48 \nl
Carina &  & dSph & -  9.4 & 0.10 & 0.51 \nl
Leo A & DDO 69 & Ir V & -11.5 & 0.69 & 0.88 \nl
Leo I & Regulus & dSph & -11.9 & 0.25 & 0.61 \nl
Sextans &  & dSph & -  9.5 & 0.09 & 0.51 \nl
Leo II & DDO 93 & dSph & -10.1 & 0.21 & 0.57 \nl
Ursa Minor & DDO 199 & dSph & -  8.9 & 0.06 & 0.43 \nl
Draco & DDO 208 & dSph & -  8.6 & 0.08 & 0.43 \nl
Milky Way & Galaxy & S(B)bc I-II & -20.9: & 0.01 & 0.46 \nl
Sagittarius &  & dSph(t) & -13.8:: & 0.03 & 0.46 \nl
SagDIG &  & Ir V & -12.0 & 1.18 & 1.29 \nl
NGC 6822 &  & Ir IV-V & -16.0 & 0.50 & 0.67 \nl
Aquarius & DDO 210 & V & -10.9 & 0.95 & 0.95 \nl
Tucanae &  & dSph & -  9.6 & 0.87 & 1.10 \nl
Cassiopeia & And VII & dSph & -12.0 & 0.69 & 0.29 \nl
Pegasus & DDO 216 & Ir V & -12.3 & 0.76 & 0.44 \nl
Pegasus II & And VI & dSph & -11.3 & 0.78 & 0.38 \nl
\enddata
\tablenotetext{a}{Galaxies listed in order of increasing Right Ascension}
\end{deluxetable}

\begin{figure}[h]
\centerline{\epsfysize=7in%
\epsffile{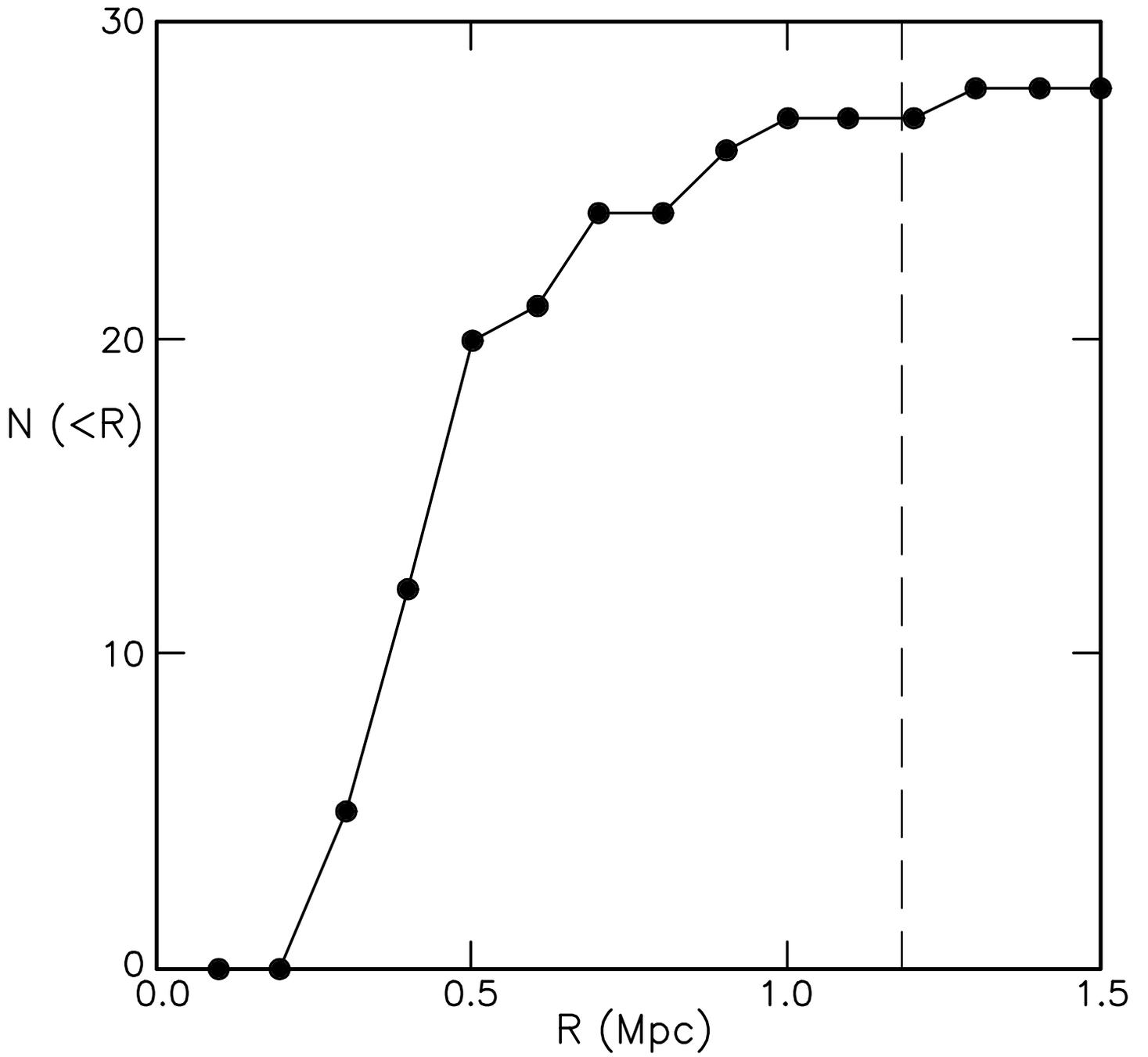}}
\caption{Number of Local Group members within a distance R 
of the barycenter of the Local Group.  To minimize the effects 
of observational selection galaxies fainter than M$_{\rm V} = -10.0$ 
were excluded.  The dashed vertical line is the dynamically 
determined radius of the Local Group zero-velocity surface.}
\end{figure}

\end{document}